\documentclass[]{jd04}    
\usepackage[]{graphicx}

\Title[Spectra of halo stars within 3550--5000\,\AA\AA{}]
     {High resolution spectroscopy of halo stars within the spectral
     region 3550--5000\,\AA\AA}

\Author{Klochkova$^1$, Valentina}
\Author{Ermakov$^1$, Sergey}
\Author{Panchuk$^1$, Vladimir}
\Author{Zhao$^2$, Gang}

\Address{$^1$Special Astrophysical Observatory, Nizhnij Arkhyz, Russia \\
         $^2$National astronomical observatories of CAS,  Datun Road, Beijing, China\\}{}

\ShortAuthors{Klochkova et al.}


\begin{document}
\maketitle

\begin{abstract}

\centerline{\em High resolution echelle spectroscopy in ground-based UV}

\centerline{\em is a powerful facility for study of chemical evolution
             of early Galaxy.}

\medskip

An atlas of high resolution (R\,=\,60000) in the poor studied wavelength
range 3550--5000\,\AA\AA{} for 4 metal-deficient stars in the interval of
metallicity $-3.0 < [Fe/H] < -0.6$, effective temperature 4750 $< T_{eff}
<$ 5900\,K, surface gravity $1.6 < \lg{g} < 5.0$ is produced. Details of
the method of producing a spectral atlas, line identifications, stellar
atmospheric parameters determination are described. Based on these
spectral data, we determined model atmosphere parameters and calculated
abundances of 25 chemical elements.

\end{abstract}

\section{Introduction}

During last decades we see a large growth of scientific interest in study
of the old extremely metal-poor stellar population. Search for stars of
first generations in Galaxy is a key program at largest telescopes in world
since chemical evolution of Galaxy in whole is imprinted in chemical
abundances pattern in atmospheres of stars of following generations.
Reconstruction of chemical evolution of Galaxy requires an enormous number
of highly qualitative spectra of metal deficient stars. We undertake to
high resolution spectroscopy of old stars at the 6\,m telescope of the
Special Astrophysical Observatory RAS. Below we present some results.

\section{Spectral observations and atlas producing}

We have selected a sample of metal-poor stars to obtain their evolution
status and detailed chemical composition. Some characteristics of 4 first 
stars and model parameters adopted are presented in Table\,\ref{model}.
The spectra were obtained using the echelle spectrograph NES (Panchuk et
al., 2007) permanently mounted in a Nasmyth focus. NES provides a spectral
resolving power of R$\ge$60000 in the range 3200--10000\,\AA. A
possibility to observe in UV was given thanks to creation of an echelle
spectrograph NES with a camera from fused silica. NES works in combination
with a CCD 2048\,x\,2048 pixels having a high sensitivity in a blue
spectral range.

\begin{figure}[hbtp]
\centerline{\includegraphics[angle=0,width=13cm, height=4.8cm, bb=0 70 260 260,clip=]{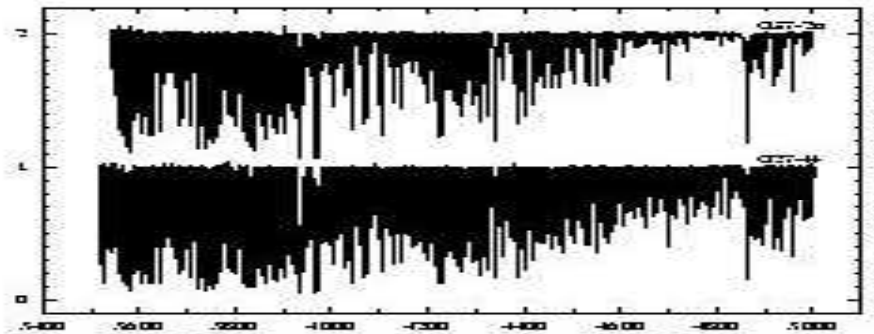}}
\caption{Comparison of the spectra of two stars with very distinguished metallicity
         (G37--26, [Fe/H]=$-2.04$, and G27--44, [Fe/H=$-0.60$ within the full spectral
	 region registered.}
\label{Full}
\end{figure}

The 2D spectra were reduced (applying standard procedures of bias
subtraction, scattered light and cosmic ray trace removal, and order
extraction) using the context ECHELLE under MIDAS (version 01FEB). To
process spectra obtained with an image slicer the context ECHELLE was
modernized (Yushkin \& Klochkova, 2004). The signal-to-noise ratio for all
the spectra shown in this atlas is higher than 200. Combined with the
spectral resolving power, that allowed us not only to detect rather weak
lines but also to study their profiles. To illustrate an effect of
difference in metallicity, full spectra of 2 stars with significantly
distinqished metallicity are presented in Fig.\,\ref{Full}. Such a scale
illustrates very well a role of blanketting effect. Large details, Balmer
lines (especially H$\gamma$ and H$\beta$), Ca{\sc ii} doublet, are clearly
seen.

\begin{table}[htp!]
\begin{center}
\caption{Basic data and  model atmospheric parameters of stars studied}
\begin{tabular}{l|crccccc}
\hline
Star     &$\rm V$&  $\pi$, {\it mas}& $T_{eff}$, {\it K} & $\lg{g}$ & [Fe/H]&  $\xi_t$, {\it km/s} \\
\hline
G27-44    &  7.411  & 23.66 & 4975 & 4.35 & $-0.60$ &  1.0 \\
HD\,188510&  8.832  & 25.32 & 5475 & 5.00 & $-1.52$ &  0.6 \\
G37-26    &  8.056  & 25.85 & 5900 & 4.50 & $-2.04$ &  0.7 \\
HD\,115444&  8.975  & 3.55  & 4750 & 1.60 & $-2.91$ &  1.7 \\
\hline
\end{tabular}
\label{model}
\end{center}
\end{table}

\begin{figure}[hbtp]
\centerline{\includegraphics[width=12cm,height=6.5cm, bb=150 90 750 550,clip=]{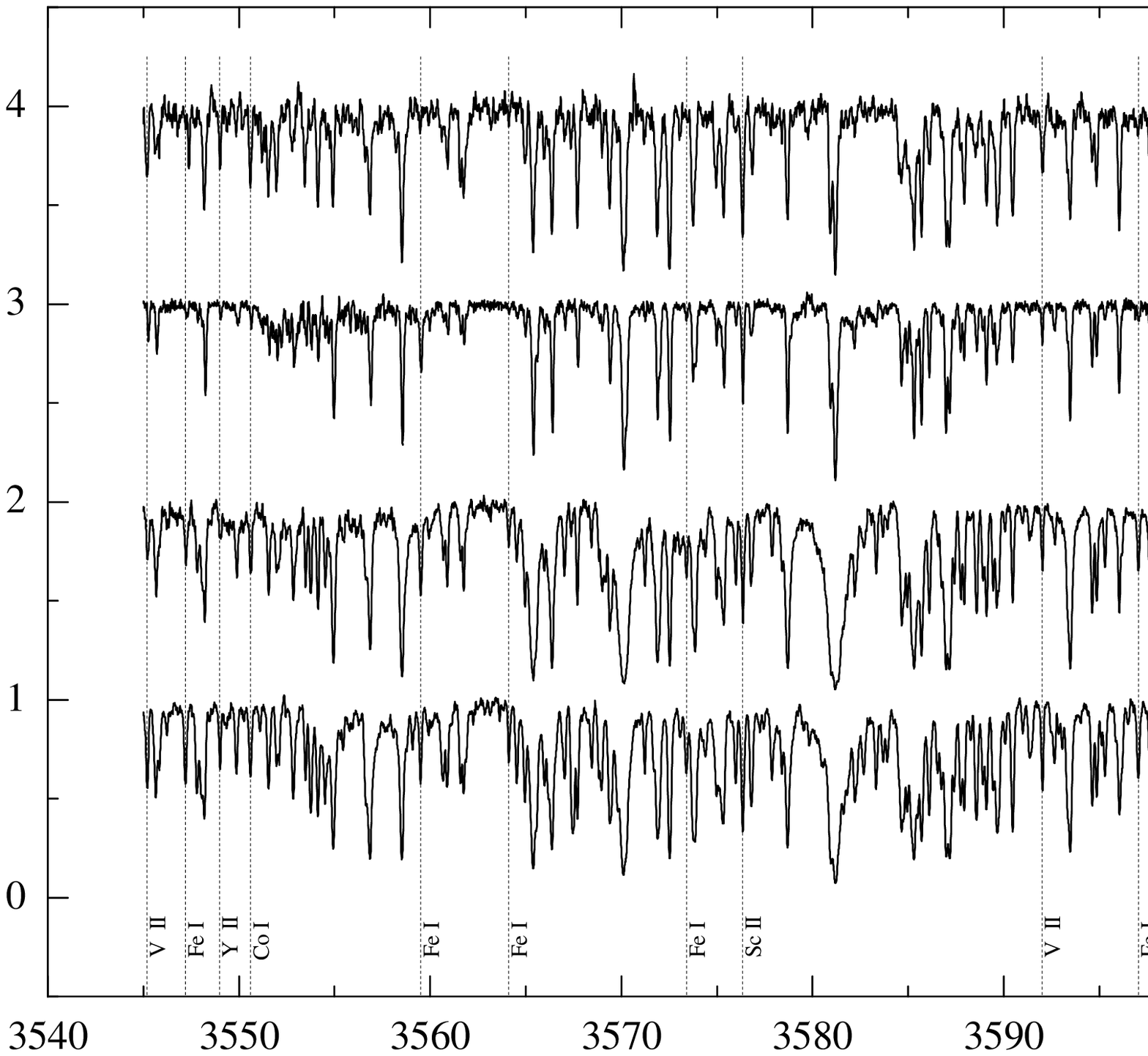}}
\caption{One fragment of the atlas.}
\label{UV-fragment}
\end{figure}

\begin{figure}[hbtp]
\centerline{\includegraphics[angle=0,width=11.0cm, height=4.0cm, bb=28 30 264 204,clip=]{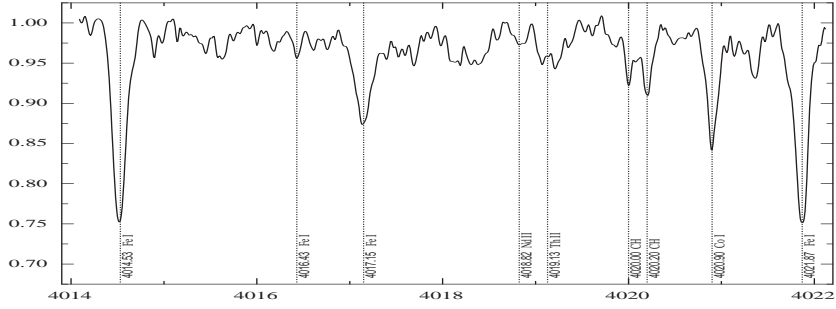}}
\caption{The fragment of the spectrum of a metal-poor star including the
         Nd{\sc ii} 4018 and Th{\sc ii} 4019\,\AA{} lines.}
\label{ThNd}
\end{figure}

We made the atlas as intensities, normalized to the continuum, versus
laboratory wavelengths in the range 3550--5000\,\AA{}. The atlas includes
29 spectral fragments approximately 60\,\AA\ in width. As an example, the
region $\lambda$\,3550--3600\,\AA{} is shown in Fig.\,\ref{UV-fragment}.
For orientation in the identification some principal details used for
chemical composition calculation are marked. The lines in the spectra were
identified using the data from the VALD (Piskunov et al., 1995). The
initial list of lines includes about 8100 lines. Based on the solar
spectrum, about 860 unblended lines  were selected. Quality of spectra
in near UV-region permits us to begin the task of cosmochronology (see
Th and Nd lines in Fig.\,\ref{ThNd}).

\section{Model parameters and elemental abundances}

The effective temperature $T_{eff}$ was derived using the Str\/{o}mgren indices ($
b-y $, $c_1$) and the calibration based on the method of infrared flux
(Alonso et al., 1996).  The metallicity values needed for the first iteration of
$T_{eff}$ determination were used from publications, but in the following
iterations our spectrocopic $\rm [Fe/H]$ values were used. In such a
procedure for different stars from 100 to 230 Fe{\sc i} lines were taken into
account. Besides, to control the $T_{eff}$ values we added the generally
used spectroscopic way $T_{eff}$ determination -- a forcing of independence
of $\lg\epsilon(Fe)$ on low level excitation potential. A surface gravity
were calculated based on  known relations:
				
$$\lg{\frac{g}{g_{\odot}}} = \lg{\frac{\mathcal{M}}{\mathcal{M_{\odot}}}} +
     4\lg{\frac{T_{eff}}{T_{eff_{\odot}}}}+0.4(M_{bol} - M_{bol,\odot}),$$ 
where: $M_{bol} = V + BC + 5\lg{\pi}$+5, $\mathcal{M}$ -- a mass of a star,
     $M_{bol}$ -- bolometric luminosity,
     $V$ -- visual magnitude, 
     $BC$ -- bolometric correction, 
     $\pi$ -- stellar parallax.

The Hipparcos parallaxes were attracted for calculations. The stellar
masses were determined using the evolution tracks by Vandenberg et al.
(2000) calculated with a step in metallicity ${\rm \approx 0.1 dex}$.
Bolometric corrections were calculated using the calibration formula by
Balona (1994). The microturbulent velocity $\xi_t$ was determined forcing
the indepedence of neutral iron abundance on equivalent width
W$_{\lambda}$ of the line. The programme WIDTH9 and the Kurucz's grid of
the atmospheric models were used for chemical abundances calculation.

\begin{table}[htp!]
\begin{center}
\caption{Relative abundances of chemical elements [X/Fe]}
\begin{tabular}{lrrrr}
\hline
Species &\multicolumn{4}{c}{[X/Fe]=${(\lg\epsilon(X) - \lg\epsilon(Fe)})_* -
	({\lg\epsilon(X) - \lg\epsilon(Fe)})_{\odot}$}  \\
\cline{2-5}
&  G27-44 &HD\,188510& G37-26 &HD\,155444 \\
\hline
Na1 &  $-0.12$ &  $     $ &$       $ & $ +0.30$ \\                     
Mg1 &  $+0.16$ &  $+0.27$ &$  +0.55$ & $ +0.77$ \\                     
Al1 &  $-0.65$ &  $-0.93$ &$  -0.72$ & $ -0.12$ \\		      
Si1 &  $-0.24$ &  $-0.27$ &$  +0.11$ & $ +0.53$ \\		      
Ca1 &  $+0.31$ &  $+0.43$ &$  +0.53$ & $ +0.48$ \\                    
Sc2 &  $+0.37$ &  $+0.25$ &$  +0.19$ & $ +0.23$ \\		      
Ti1 &  $+0.06$ &  $+0.19$ &$  +0.31$ & $ +0.35$ \\                    
Ti2 &  $+0.20$ &  $+0.32$ &$  +0.37$ & $ +0.41$ \\		      
V1  &  $+0.03$ &  $+0.06$ &$  +0.14$ & $ +0.16$ \\		      
V2  &  $+0.32$ &  $+0.25$ &$  +0.13$ & $ +0.17$ \\
Cr1 &  $-0.01$ &  $+0.04$ &$  -0.05$ & $ -0.23$ \\
Cr2 &  $+0.20$ &  $+0.18$ &$  +0.16$ & $ +0.23$ \\
Mn1 &  $-0.10$ &  $-0.27$ &$  -0.33$ & $ -0.47$ \\
Fe1 &  $+0.00$ &  $-0.00$ &$  +0.01$ & $ -0.00$ \\
Fe2 &  $-0.00$ &  $+0.00$ &$  -0.01$ & $ +0.00$ \\
Co1 &  $+0.13$ &  $+0.19$ &$  +0.27$ & $ +0.36$ \\
Ni1 &  $-0.09$ &  $-0.04$ &$  -0.09$ & $ -0.01$ \\
Zn1 &  $-0.08$ &  $+0.09$ &$  +0.10$ & $ +0.26$ \\
Sr2 &  $-0.03$ &  $-0.35$ &$  +0.16$ & $ +0.07$ \\
Y2  &  $+0.10$ &  $-0.00$ &$  +0.01$ & $ +0.04$ \\
Zr2 &  $+0.06$ &  $+0.43$ &$  +0.36$ & $ +0.35$ \\
Ba2 &  $+0.21$ &  $+0.10$ &$  +0.26$ & $ +0.72$ \\
La2 &  $+0.23$ &  $+0.30$ &$       $ & $ +0.33$ \\
Ce2 &  $+0.25$ &  $+0.38$ &$  +0.37$ & $ +0.35$ \\
Nd2 &  $+0.45$ &  $+0.73$ &$  +0.48$ & $ +0.60$ \\
Sm2 &  $+0.31$ &  $+0.82$ &$  +1.41$ & $ +0.77$ \\
Eu2 &  $+0.66$ &  $+0.76$ &$  +0.14$ & $ +1.26$ \\
Gd2 &  $+0.30$ &  $+1.49$ &$       $ & $ +0.96$ \\
Dy2 &  $+0.31$ &  $     $ &$  +0.57$ & $ +0.74$ \\
\hline 
\end{tabular}
\label{relative}
\end{center}
\end{table}

From comparison our results for the star HD\,115444 with data by Westin et
al. (2000) for overlapping elements we obtained systematic deviation about
0.1\,dex which is caused by using of different systems of oscillator
strengths and models parameters adopted. For most of chemical elements we
obtained typical for their metallicity chemical abundances picture. For
example, $\alpha$-process elements Mg, Ca, Ti are overabundant for stars
of low metallicity.

\section*{Summary}

For the first time an unique atlas of F--K-stars spectra of very low
metallicities is produced. An atlas is produced for the range
3550--5000\,\AA\AA{} with a high spectral resolution R\,=\,60000. The lines
identification was performed by the models atmospheres method. More
detailed results will be published in a forthcoming paper by Klochkova et
al. (2006). The atlas in whole, W$_{\lambda}$, atomic data  and
abundances calculated are available by web access to
$http://www.sao.ru/hq/ssl/Atlas-UV/Atlas-UV.html$. A comparison of the spectra of stars
may help in a search for new spectroscopic criteria to distinguish metal
deficient stars. In this connection several spectroscopic features are
particularly interesting. Such spectra could be useful for studies both of
the chemical abundances of separate stars and determination their
evolution stages. The atlas provides a spectral library for
population synthesis. 
Besides a large number of spectral lines sufficient for calculation of
numerous elements abundances, a short wavelengths range has another
advantage in task of chemical composition determination. For temperature
and pressure values typical for subdwarf atmospheres a coefficient of
extintion in continuum near 4000\,\AA{} is lower about 0.2\,dex than in red
part of spectra. It means that continuum extintion arises deeper in
atmosphere than that in red range (near 6000\,\AA{}). Therefore, weak
short wavelengths lines are formed in average closer to photosphere than
longer wavelengths lines. It permits to hope that a model description of
short wavelengths line is more accurate than the longer wavelengths lines.

\medskip
\noindent
{\it Acknowledgements:}

Our work was supported by the Russian Foundation for Basic Research
(project 05--07--90087\,b) and the Russian Federal program ``Observational
manifestations of evolution of chemical abundances of stars and Galaxy''.

\end{document}